# Maximum entropy distribution of stock price fluctuations

Rosario Bartiromo[a]

Istituto di Struttura della Materia del CNR, via Fosso del Cavaliere 100, 00133 Roma,

and

Dipartimento di Fisica, Università di Roma Tre, via della Vasca Navale 84, 00146 Roma,

Italy

In this paper we propose to use the principle of absence of arbitrage opportunities in its entropic interpretation to obtain the distribution of stock price fluctuations by maximizing its information entropy. We show that this approach leads to a physical description of the underlying dynamics as a random walk characterized by a stochastic diffusion coefficient and constrained to a given value of the expected volatility, taking in this way into account the information provided by the existence of an option market. The model is validated by a comprehensive comparison with observed distributions of both price return and diffusion coefficient. Expected volatility is the only parameter in the model and can be obtained by analysing option prices. We give an analytic formulation of the probability density function for price returns which can be used to extract expected volatility from stock option data

89.65.Gh, 05.45.Tp, 89.70.Cf

[a] e-mail: bartiromo@fis.uniroma3.it

**Introduction**

The behaviour of stock prices has attracted considerable attention by the physics community in recent years[1,2,3]. More than by its practical interest for asset allocation and risk management, physicists are particularly intrigued by two features of the distribution of price returns $x(t) = \log(p(t)/p(t-\Delta t))$, where $\Delta t$ is the observation time and p(t) the price value at time t. First, it is markedly non-Gaussian, with fat tails enhancing the probability of large fluctuations by many orders of magnitude[4]. Second, when returns are normalized to their standard deviation, different stocks and different markets show very similar probability distributions or in other words the return distribution scales to a common shape[4,5], as we will document again in the following. Moreover this universal distribution shows a mild dependence upon the observation time $\Delta t$ as long as it remains shorter than 16 trading days[6]. The challenge to find a simple mechanism to explain this neat behaviour has not yet been met.

A description of the observed phenomenology in terms of a stable Levy process was first put forward by Mandelbrot[7] and analysed in great details in subsequent years by many authors over a large body of market data. It requires that the cumulative distribution of large returns is well described by a power law, in reasonable agreement with observations[5,6]. However the exponent in this power law should be higher than 2 to be compatible with a diverging variance as required by the Levy hypothesis, a feature which is not confirmed by observations. We note in passing that a diverging variance of price returns would make it very difficult, if not impossible, the task of quantifying financial risks.

An alternative approach to account for the existence of fat tails assumes that the price distribution is not stationary but depends on market conditions[8,9]. More specifically, a Gaussian distribution is expected for a stationary market, since the stock price would depend on the stochastic behaviour of a large number of market agents and, in force of the central limit theorem, this would produce a Gaussian. The standard deviation of this elementary distribution would then quantify the risk of keeping the stock in portfolio and thus should in principle be a time dependent quantity. In the following of this paper we will refer to the standard deviation of price return distribution as the price volatility.

These considerations have motivated the development of stochastic volatility models[10,11] which assume a parametric form for an auxiliary process which regulates the instantaneous value for the width of the elementary Gaussian distribution. In econometrics autoregressive models, known with the acronyms ARCH[12] and its numerous modifications[13,14,15,16], were developed to forecast volatilities and, depending on the particular version adopted, have been shown able to reproduce many of its observed characteristics such as long range memory[17] and leverage effects[18]. Optimal values for the model parameters can be obtained by fitting the observed distribution of price returns[19] or, as recently done by Gerig et al.[20], by fitting the observed distribution of daily volatility. Although these models are able to follow the time evolution of volatility and can be usefully adopted for a variety of tasks related to risk assessment and derivative contract pricing, they remain phenomenological in nature and do not tell us very much about the mechanism which leads to the observed price fluctuations.

In this paper we focus on the observed scaling and universality of price fluctuation distributions. We will be lead by a physics based description of market mechanisms to a double stochastic model suitable to represent price fluctuations. This will enable us to compute the entropy of the process. Subsequently we will assume that the mere observation of price fluctuations does not provide any new information, as implied by the principle of absence of arbitrage opportunities. This allows to maximize the process entropy which we exploit to find the marginal volatility distribution and finally the price fluctuation distribution. The final result depends critically on the public information available to market participants and we will use market data to discriminate among different plausible options. This analysis will show that the best physical description of price fluctuation dynamics is a random walk characterized by a stochastic diffusion coefficient and constrained to a given value of the expected volatility, taking in this way into account the information provided by the existence of an option market.

The paper is structured as it follows: we will first describe in section II how a stock market works and we will introduce the concept of an ideally liquid stock to simplify its physical description. The basic formalism of a double stochastic model will be given in section III while in section IV we will calculate its information entropy and discuss the procedure and the constraints used to maximize its value. An extensive comparison with market data for daily price returns will be presented in section V and in the following section VI use will be made of tick data to evaluate diffusion coefficients to compare with the predictions of our models. Section VII will be devoted to the issue of stationarity and its influence on the tails of the distribution. In the last section we will discuss our conclusions.

**II Stock market operation**

Stock prices are set in a continuous double auction[21] whereby trader orders that are not immediately executed at the best available price are stored in the order book in two queues, the bids which are buy orders and the asks which are sell orders. Prices are quantized and the minimum price variation, the tick τ, can be different from stock to stock. Therefore a double auction is not characterized by a single price but rather by the best bid, which is the price a potential seller would get, and the best ask, which is the price a potential buyer would need to pay in order to get the shares. In a previous work[22] we have shown that to describe the stochastic dynamics of stock prices as a continuous time random walk the system has to be considered in a waiting state as long as both the best bid and the best ask prices remain constant. A simple algorithm can then easily identify a transition from the price sequence in a transaction database when the mid-price between the best ask and the best bid changes and quantify the step amplitude as the corresponding variation (usually corresponding to one tick for liquid stocks).

To simplify the understanding of our analysis we introduce the concept of an ideally liquid stock[23]. We shall say that a stock is ideally liquid when all levels in its order book are always filled and its mid-price can only change by the tick size. Then the stochastic dynamics of an ideally liquid stock is that of a binomial random walk characterized by the number N of transitions recorded during the observation time Δt and the probability η to observe an upward transition. We take η equal to ½ neglecting the small variations allowed by the efficient market hypothesis. Under these assumptions the price return should perform a random walk with diffusion coefficient given by $D = \frac{\tau^2}{p^2} f$ where $f$ is the transition frequency. Price return volatility on the scale Δt is therefore given by $w = \sqrt{D\Delta t} = \sqrt{N\left(\frac{\tau}{p}\right)^2}$ provided the observation time Δt is long enough to allow a number of transitions higher then about 10. If w does not change with time the return distribution is normal with zero mean and variance equal to $w^2$. However the transition frequency is observed to change both on a day to day and on an intraday basis, therefore both volatility and diffusion must be time dependent quantities in this approach.

This analysis of market mechanisms suggests that traders may adapt to changing risk perception by varying the rate at which they trade and therefore mainly by means of the transition frequency f. In extreme situations they could also open the bid-ask spread to more than one tick but in both circumstances it appears that from a physicist standpoint the diffusion coefficient would be the dynamical variable that controls the volatility value w.

**III Double stochastic model**

Following the stochastic volatility approach[24] we assume that observed price returns x can be considered as the output of a process whose parameters are functions of another stochastic variable set by a different process independent from the previous one. In the following we will refer to the first of these processes as the slave process while the other will be referred to as the master process.

Without loss of generality the probability distribution function $f_w(x)$ of the slave process can be assumed of the form

$$f_w(x) = \frac{g(x/w)}{wN_0} \qquad (1)$$

with $N_0 = \int_{-\infty}^{+\infty} g(z)dz$ and w, a positive quantity, being the scale of the distribution. As discussed above, when dealing with stock price returns it is appropriate to take $f_w(x)$ as Gaussian and to choose for w the value of the standard deviation of the distribution function. This implies that the distribution $g(z)/N_0$ is normal with unitary variance and therefore $g(z) = \exp(-z^2/2)$

In the double stochastic process we need to consider here, this scale w is distributed according to a stationary function F(w) and the observed distribution of the quantity x will be given by

$$f(x) = \frac{1}{N_0} \int_0^{+\infty} F(w) \frac{g(x/w)}{w} dw \qquad (2)$$

Should volatility depend either explicitly or implicitly upon time a further averaging would be necessary but we will not pursue this approach in details in this paper.

We note at this point that, if the distribution function of the slave process is known, it is possible to obtain all the moments of F(w) from the moments of the observed stochastic variable x. Indeed the following relation holds: $\langle |x|^\alpha \rangle = \frac{N_\alpha}{N_0} \langle w^\alpha \rangle$ with

$$N_\alpha = \int_{-\infty}^{+\infty} |z|^\alpha g(z) dz.$$

If we define $\tilde{x} = \frac{x}{l}$ and $\tilde{w} = \frac{w}{w_0}$ thereby normalizing x to a scale l and w to a scale $w_0$ the scaled distribution for x becomes

$$f(\tilde{x}) = \frac{1}{N_0} \frac{l}{w_0} \int_0^{+\infty} \frac{F(\tilde{w})}{\tilde{w}} g(\frac{\tilde{x}}{\tilde{w}} \frac{l}{w_0}) d\tilde{w} \qquad (3)$$

In the analysis of stock prices it is customary to set $l = \langle |x|^\alpha \rangle^{\frac{1}{\alpha}}$, often with α=2; therefore

$$\frac{l}{w_0} = \left( \frac{N_\alpha}{N_0} \langle \tilde{w}^\alpha \rangle \right)^{\frac{1}{\alpha}} \quad (4)$$

Equations (3) and (4) show that the scaled distribution will be universal for any α as long as $F(\tilde{w})$ is universal, i.e. when the distribution of the master process has only one scale parameter $w_0$ which can depend on the specific stock being analysed. This observation will be exploited in the following section for the evaluation of the expression for F(w).

**IV Information entropy**

A sound approach to a rational evaluation of financial assets is the principle of absence of arbitrage opportunity[25]. It implies that very little information can be gained from the observation of price evolution and leads quite naturally to the idea that the distribution of price returns should maximize Shannon entropy[26].

A role for information entropy in econometrics beyond its use as a mere tool of statistical inference was first postulated by Gulko[27] who conjectured that entropy maximization could be an intrinsic feature of competitive markets and efficient pricing. Its entropic market hypothesis implies the absence of arbitrage opportunities. He also analysed the microstructure of an efficient market showing that it should operate close to the maximum entropy distribution.

Consequently we will require that the information entropy of our double stochastic process should be maximized once all the publicly available information is taken into account. We will show in the following how this allows obtaining the volatility distribution without assuming any specific model for its stochastic behaviour.

To this purpose we will assume that the scale w of the slave process depends deterministically on the value of a stochastic variable y, which we identify as the dynamic variable. Denoting its probability distribution function by F(y), the information entropy of our double stochastic process becomes, following Shannon[18]

$$S = -\int_{-\infty}^{+\infty} dy \int_{-\infty}^{+\infty} F(y) f_{w(y)}(x) ln[F(y) f_{w(y)}(x)] dx = S_g + S_F$$

with $S_g = -\int_{-\infty}^{+\infty} \frac{g(z)}{N_0} \ln\left[\frac{g(z)}{N_0}\right] dz$ and $S_F = -\int_{-\infty}^{+\infty} F(y) \ln\left[\frac{F(y)}{w(y)}\right] dy$

Therefore the contributions to the total entropy of the two distributions $g(z)/N_0$ and $F(y)$ are additive and we can maximize them separately. Maximizing $S_g$ under the additional constraint of unitary variance yields $g(z) = \exp(-z^2/2)$ as required in the previous section. The distribution F(y) can be obtained maximizing $S_F$ following the procedure first introduced by Jaynes[28]. If information is available that can be cast in terms of the average values of some functions $Q_s(y)$ then the distribution function, conditional on the available information, becomes

$$F(y) = \frac{w(y) \exp[-\sum_s \lambda_s Q_s(y)]}{\int_{-\infty}^{+\infty} w(y) \exp[-\sum_s \lambda_s Q_s(y)] dy}$$

where $\lambda_s$ are Lagrange multipliers to be determined from the known average value of the quantities $Q_s(y)$. This yields for the original variable w

$$F(w) = \frac{wM(w) \exp[-\sum_s \lambda_s Q_s(w)]}{\int_0^{+\infty} wM(w) \exp[-\sum_s \lambda_s Q_s(w)] dw} \qquad (6)$$

where $M(w) = \dfrac{1}{dw/dy}$. This result which could have been obtained by maximizing the Kullbach entropy[29] with measure M(w) but we have chosen here to show explicitly the meaning and the origin of the function M.

In a language more familiar to physicists we recall that M(w) represents the density of states and that for a dynamic variable the corresponding density of states is a constant. In our model of ideally liquid stocks the diffusion coefficient D and the variance $w^2$ are both proportional to the number N of transition occurring in the observation time $\Delta t$. Since in this model N uniquely identifies a state of the stochastic system, the variance, or equivalently the diffusion coefficient, is the dynamical variable in our problem. We therefore take $M(w) = w^{\alpha_M}$ with $\alpha_M=1$. For sake of comparison in the following we will also consider the case $\alpha_M=0$ to test whether the volatility itself could be a viable option for the dynamic variable.

To proceed further in our construction of the function F(w) we have to find out which information is available to use in equation (6). We start by noting that in the market each trader builds his own strategy based on the information, both private and public, available to him. The observed return distribution is the aggregate result of the actions of all traders. The only information that is likely to survive in the aggregation process is public information available to all of them and worth to be used in their trading strategies. The principle of absence of arbitrage opportunities implies that no public information is available that could be useful to predict the direction of price fluctuations.

On the contrary information about the expected value of their amplitudes, that is price volatility, is readily and publicly available when an efficient option market exists. Indeed option markets are volatility markets and, when they are efficient, they price it at its expected value which is therefore publicly available. Of course this does not tell anything about the volatility observed in a particular trading day but gives useful information on the average volatility expected by the market over the time interval until the option expiration date. In a recent paper[23] we described an analysis of the information entropy of volatility distributions for stocks traded on the Italian stock market in Milan and we documented the existence of a shared knowledge of its expected value.

We now recall that, since we observe scaling in the distribution of returns, we are looking for a volatility distribution depending on just one scale parameter. This indicates that conditioning on the expected value of volatility should be sufficient to obtain a useful distribution. We therefore use $Q(w)=w^\delta$ with $\delta=1$. Again for sake of comparison, since variance is the dynamic variable, in the following we will also consider the case $\delta=2$ to test whether conditioning on variance expectation could offer an alternative.

Therefore we will write the volatility distribution function as

$$F(w) = \frac{w^{\alpha_M+1}\exp[-\lambda w^\delta]}{Z(\lambda)} \qquad (7)$$

with $Z(\lambda) = \int_0^{+\infty} w^{\alpha_M+1}\exp[-\lambda w^\delta]dw = \dfrac{\Gamma\left(\dfrac{2+\alpha_M}{\delta}\right)}{\delta \lambda^{\frac{2+\alpha_M}{\delta}}}$ where $\Gamma$ represents the Gamma function.

It is worth to remark at this point that F(w) as given above represents the maximum entropy distribution conditional on a given value for the expectation value $<w^\delta>$ which determines the value of $\lambda$. We can use it as a reasonable approximation to the marginal distribution we need in the data analysis only under the additional hypothesis that $<w^\delta>$ does not change very much over the time span covered by our dataset. With this caveat and since $<w^\delta> = -\frac{\partial}{\partial \lambda} \ln Z(\lambda) = \frac{2+\alpha_M}{\delta} \frac{1}{\lambda}$ we obtain the characteristic length of the w distribution as $w_0 = \left(\frac{\delta}{2+\alpha_M}\right)^{\frac{1}{\delta}} <w^\delta>^{\frac{1}{\delta}}$ and in the normalized units $\tilde{w} = \frac{w}{w_0}$ we can write

$$F(\tilde{w}) = \frac{\delta}{\Gamma\left(\frac{2+\alpha_M}{\delta}\right)} \tilde{w}^{\alpha_M+1} \exp[-\tilde{w}^\delta] \qquad (8)$$

We note in passing that this expression, with $\delta=2$ and suitable choices for $\alpha_M$, would reproduce the stationary distribution of stochastic volatility models based on the Heston process or the exponential Ornstein-Uhlenbeck process[11]

The generic momentum of w is readily obtained from equation (8) as

$$<w^\alpha> = \frac{\Gamma\left(\frac{2+\alpha_M+\alpha}{\delta}\right)}{\Gamma\left(\frac{2+\alpha_M}{\delta}\right)} w_0^\alpha \qquad (9)$$

Defining $x_\alpha = \frac{x}{<x^\alpha>^{\frac{1}{\alpha}}}$ and using relations (4), (8) and (9) equation (3) yields:

$$f(x_\alpha) = \frac{1}{N_0} \frac{\delta}{\Gamma\left(\frac{2+\alpha_M}{\delta}\right)} \left[\frac{N_\alpha}{N_0} \frac{\Gamma\left(\frac{2+\alpha_M+\alpha}{\delta}\right)}{\Gamma\left(\frac{2+\alpha_M}{\delta}\right)}\right]^{\frac{1}{\alpha}} \int_0^{+\infty} g\left(\frac{x_\alpha}{\tilde{w}}\left[\frac{N_\alpha}{N_0} \frac{\Gamma\left(\frac{2+\alpha_M+\alpha}{\delta}\right)}{\Gamma\left(\frac{2+\alpha_M}{\delta}\right)}\right]^{\frac{1}{\alpha}}\right) \tilde{w}^{\alpha_M} \exp[-\tilde{w}^\delta] d\tilde{w}$$

In the comparison with market data we will take in the following $\alpha=2$ to be consistent with most of the published literature.

## V  Comparison with market data

The discussion in the previous section indicates that an appropriate physics based mechanism underlying the observed return distribution would be that of a random walk with a stochastic diffusion coefficient and a given average for the standard deviation. We will identify this model in the following by the notation ST11 from the value of the two coefficient $\alpha_M$ and $\delta$. For sake of comparison we will also consider an ST01 model where the volatility itself is the dynamic variable thereby taking $\alpha_M = 0$ with $\delta=1$ and a ST12 model where we constrain to the average of the diffusion coefficient, the dynamical variable, i.e. $\alpha_M = 1$ and $\delta = 2$.

To compare these three models with market data we exploit the availability of seven datasets of stock prices collected in five different markets which are described in table I. We first concentrate on daily returns. In fig. 1 we show the cumulative distribution of absolute returns for 20 Blue Chips stocks traded at the London Stock Exchange (LSE). Data refer to a time span covering the years from 1990 to 2006 and returns have been normalized to the sample estimate of their standard deviation. They are displayed on a log-log plot to ease comparison with most of the published literature. The figure shows good evidence of the scaling behaviour described above as well as a clear deviation from a Gaussian with a distinct tail on the side of high absolute returns. Similar results are obtained for all the remaining daily datasets.

For the purpose of this analysis it is then justified to consider the data relative to each stock as a different realization of the same stochastic process. Our data analysis strategy consists in building a scaled distribution for each dataset available as the average of the scaled distribution of single stocks with an uncertainty given by the standard deviation. In fig. 2 we show this average distribution, together with a typical error bar, for each market represented in Table 1. This plot shows that the distribution of normalized return does not depend on the market except in the far tail comprising much less than 1% of the all the events recorded.

This finding allows building a weighted average of the different market distributions to compare with the models we developed before. Such a comparison is illustrated in fig. 3 which shows that a good match to market data is observed for the ST11 model that is able to reproduce about 99.9 % of the observed cumulative distribution, corresponding to a return about 5 times the standard deviation. For this level of fluctuation a Gaussian would underestimate the observed data by more than 4 orders of magnitude.

Of the remaining two models, the ST12 underestimates the data in the large return region while the ST01 overestimate the distribution already at a normalized return of 3. Similar results are obtained for the average distribution of each daily datasets shown in fig. 2. This supports the arguments we used above to identify the available information and the dynamical variable of the process.

A critical reader would note that up to this point our model is merely a description of observed data, in line with many applications of the maximum entropy approach[28]. And indeed we have used the sample average of the standard deviation to normalize the data. However the previous analysis suggests that returns should be normalized to the expected volatility value. This information could be obtained from the analysis of option prices but unfortunately such data are not easily available in the quantity needed for a comprehensive comparison. In our analysis normalization to the sample standard deviation is used to obviate to this lack of information.

To extract the value of implied volatility from option prices it is important to note that the integral in equation (2) can be obtained explicitly for all the models considered so far. In particular for the ST11 model equation (8) gives $F(w) = \frac{27}{2<w>^3} w^2 \exp[-\frac{3w}{<w>}]$ that, when plugged in (2), yields

$$f\left(\frac{x}{\langle w \rangle}\right) = \frac{3}{\pi\sqrt{2}} G_{03}^{30}\left(\frac{1}{2}\left(\frac{3x}{2\langle w \rangle}\right)^2 \bigg| 0, 1, \frac{3}{2}\right) \tag{10}$$

where $G_{pq}^{mn}\left(x \bigg| \begin{matrix} a_1,...,a_n, a_{n+1}...,a_p \\ b_1,...,b_m, b_{m+1}...,b_q \end{matrix}\right)$ is a Meijer G function.

We recall that this expression does not assume any specific stochastic dynamics for the process regulating the actual volatility value. It can be usefully exploited in place of a Gaussian to evaluate volatility implied by option prices. In particular, since it gives a fair account of the amplitude of fat tails, it should be able to provide the same implied volatility independently of the strike price as we have confirmed by the analysis of a small database of short dated option prices for the Italian MIB index.

It is worthwhile to stress at this point that the predictions of all our models remain invariant when the observation time Δt is changed. Indeed they can only depend on this parameter through the volatility scale length $w_0$ which does not enter in the expression for the scaled distribution. This result is confirmed by the analysis of 2 hours returns of Italian stocks in the last database of table I, as documented in details in fig. 4.

When considering the case of large Δt where convergence to a Gaussian is observed it is worth noticing that our analysis breaks down when the return observation time exceeds the typical time interval over which the expected volatility can be considered as constant.

**VI Diffusion coefficient analysis**

Further support to our findings is obtained from the last database of table I when using tick by tick data to evaluate the distribution of measured diffusion coefficient D. Specifically we have computed an average value of D over time intervals of 15 minutes from the number of transitions observed and the average ask-bid spread[22]. In this way we have reconstructed its probability distribution for all 13 stocks represented. The average of these distributions is shown with crosses in fig. 5 and compared with the prediction of our three models. Here again we see that the ST11 model gives an acceptable description of market data while the ST12 model, that compared not badly with the bulk of the return distribution, now is clearly ruled out by the data. On the contrary the ST01 model seems to be competitive with the ST11. In this study however we can use the expected volatility to normalize the data assuming that we can approximate it with the average measured during the previous three days. Data normalized in this way are represented by the open squares in fig. 5 and are in very good agreement with the ST11 model again.

This finding indicates that considering the expected value of the volatility constant over the whole time span covered by a dataset can only be approximately accurate. In other words the expected volatility changes over time much less than the actual volatility but still it is very likely that it does not stay really constant. We can neglect its time dependence and still describe a large part of the observations but there is no reason to assume that it cannot change over time. This has to be kept in mind specially when dealing with the tails of the distribution.

**VII Extreme events**

Indeed, although we can consider our approach successful in describing the vast majority of the available data, nevertheless for all the datasets we observe that the model somewhat underestimates the probability of large price fluctuations. This occurs for normalized returns above 5 which is the region where we also observe differences in the distributions of the different datasets, see fig. 2. It is interesting to note that datasets which cover the shortest time span (Xetra, Paris and Milan) follow the model predictions up to higher returns than those covering a larger time span (DowJones and LSE). Moreover splitting a long dataset in two parts shows somewhat improved agreement with the model again pointing to a problem of lack of stationarity.

In our model market dynamics is characterized by a broad volatility distribution that maximizes the information entropy of the observed price distribution, whose scale length is set by the volatility expected value that is the only free parameter in the model. This expected value may be structural in the sense that it may depend on the structure of the market and on the overall status of the economy but it may be rather insensitive to the details of the information flow coming from the outside world and for this reason it can remain relatively constant over long time spans. It should describe the normal operation in the market when it is able to digest in this way the vast majority of the events which can affect prices.

There can be however circumstances where the market structure is somewhat compromised. For example liquidity can suddenly dry for a sudden and unforeseen event affecting a given stock or the market as a whole, as observed in the so called flash crash at Wall Street in May 2010, or worldwide financial stability can be jeopardized by troubles at a global large financial institution, as in the case of Lehman Brothers bust in October 2008. In these conditions it is plausible that also the derivative market fails and is unable to give reliable information on the expected volatility. Therefore a single scale length is not able to describe the market behaviour and a scale free approach is needed until the market stabilizes again with a new value of the expected volatility to take into account the new situation. This kind of behaviour was documented by Kiyono el al.[30] for the USA market with an analysis of high frequency data in the time period around the crash of the Black Monday in October 1987.

An indication in this direction comes also from the comparison of the behaviour of small caps and blue chips in the Italian market. Although the corresponding datasets cover the same time span, small caps deviate from the model well before blue chips and display an enhanced probability of large fluctuations. This fits well with our hypothesis that these deviations are caused by market disruptions since it is well known that small caps, especially in a rather small market like the Italian stock exchange, are much more prone to liquidity crisis.

Definite evidence of the role of non stationarity in the origin of fat tails is obtained by using a slightly modified dataset for the stocks composing the Dow Jones index. It extends from January 1990 to April 2010 to include the recent period of strong market turbulence caused by the financial crisis. The near term volatility expected by the USA market is quantified since 1990 by the VIX index computed by the Chicago Board Option Exchange and over the time span of this last dataset the yearly volatility ranges from about 9% to more than 80% with the all time maximum reached around the time of Lehman Brothers bust. When analysed as in the previous sections, these data show pronounced tails and strong deviation from the ST11 model, see fig. 6. However when we limit our analysis only to points corresponding to a VIX value between 15% and 25%, a clear reduction of tails and a satisfactory agreement with the model up to a normalized return of 10 is observed.

**Conclusions**

To put in context our findings, we recall that financial markets are complex systems driven by the flux of information coming from the outside world and regulated by an internal dynamics. We have modelled this latter dynamics with a broad volatility distribution which, according to the principle of absence of arbitrage opportunity, maximizes the information entropy of the observed price distribution.

In physical terms we have described price returns as a random walk characterized by a stochastic diffusion coefficient and constrained by a given value of the expected volatility, taking in this way into account the dynamic of the trading process and the information provided by the existence of an option market. This model has been validated by a comprehensive comparison with observations for both return and diffusion distributions. With this work we have produced the most accurate test available of the entropic market hypothesis[27] thereby improving our confidence in its relevance to the understanding of the behaviour of market participants.

Expected volatility is the only parameter in the model and can be obtained by analysing option prices and indeed from a practical point of view the most important result we have obtained with this approach is an analytic expression for the probability density function for price returns which can be used to extract the value of expected volatility from stock option prices. This distribution, given in equation (10), is ultimately related to the entropic interpretation of the principle of absence of arbitrage opportunity and gives a fairly good account of the probability of large fluctuations. Therefore in our opinion it should be preferred to a Gaussian when dealing with the problem of pricing financial derivative contracts.

| Market | Number of stocks | Time span | Data type |
|---|---|---|---|
| LSE | 60 | 1990 - 2006 | Daily |
| DowJones | 30 | 1980 -2006 | Daily |
| Xetra | 30 | 1997-2006 | Daily |
| Paris | 40 | 2000 - 2006 | Daily |
| Milan Blue Chips | 26 | 1994 - 2006 | Daily |
| Milan Small Caps | 33 | 1994 - 2006 | Daily |
| Milan | 13 | 2001 - 2004 | Tick-data |

Table I

Description of datasets used for model validation.

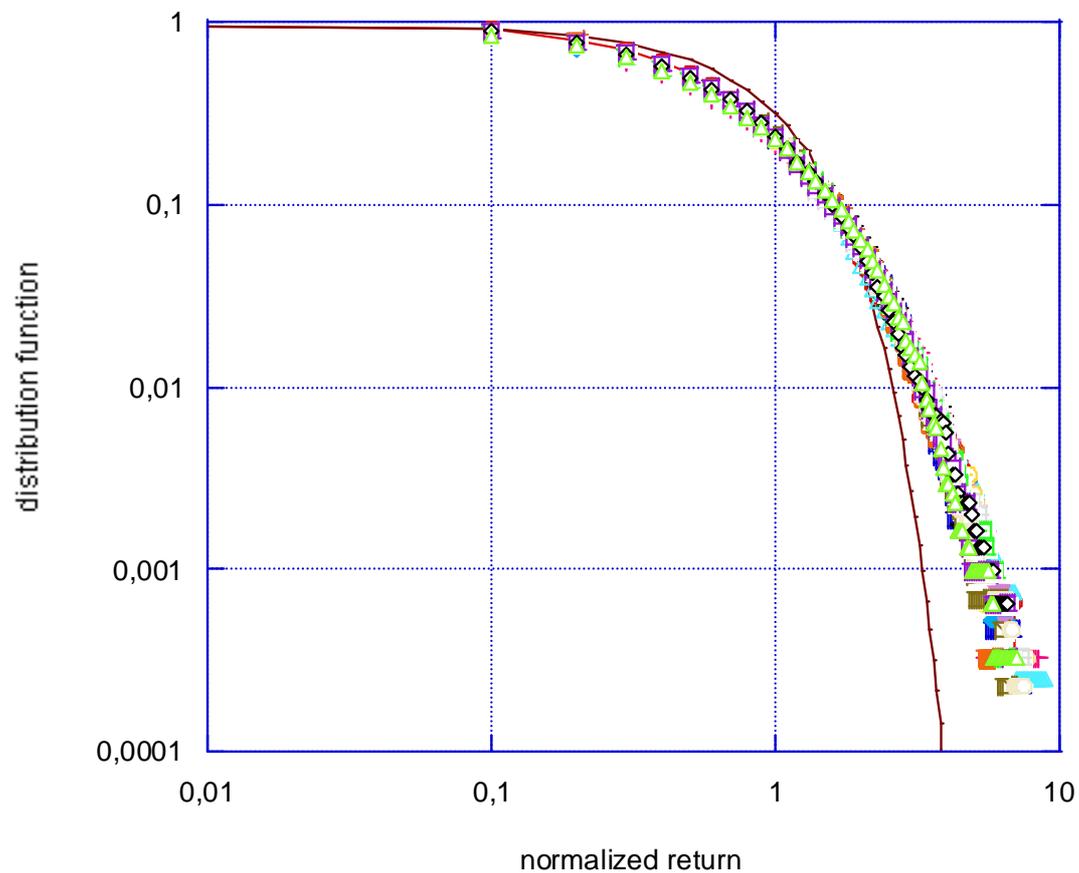

Fig. 1

The cumulative distribution function of absolute daily returns of 20 high capitalization stocks (Blue Chips) traded at the London Stock exchanges from 1990 to 2004. Data are compared to the Gaussian distribution represented by the continuous line. On the abscissa the return is normalized to the standard deviation of each database. Data are shown in a log-log plot to ease comparison with published literature.

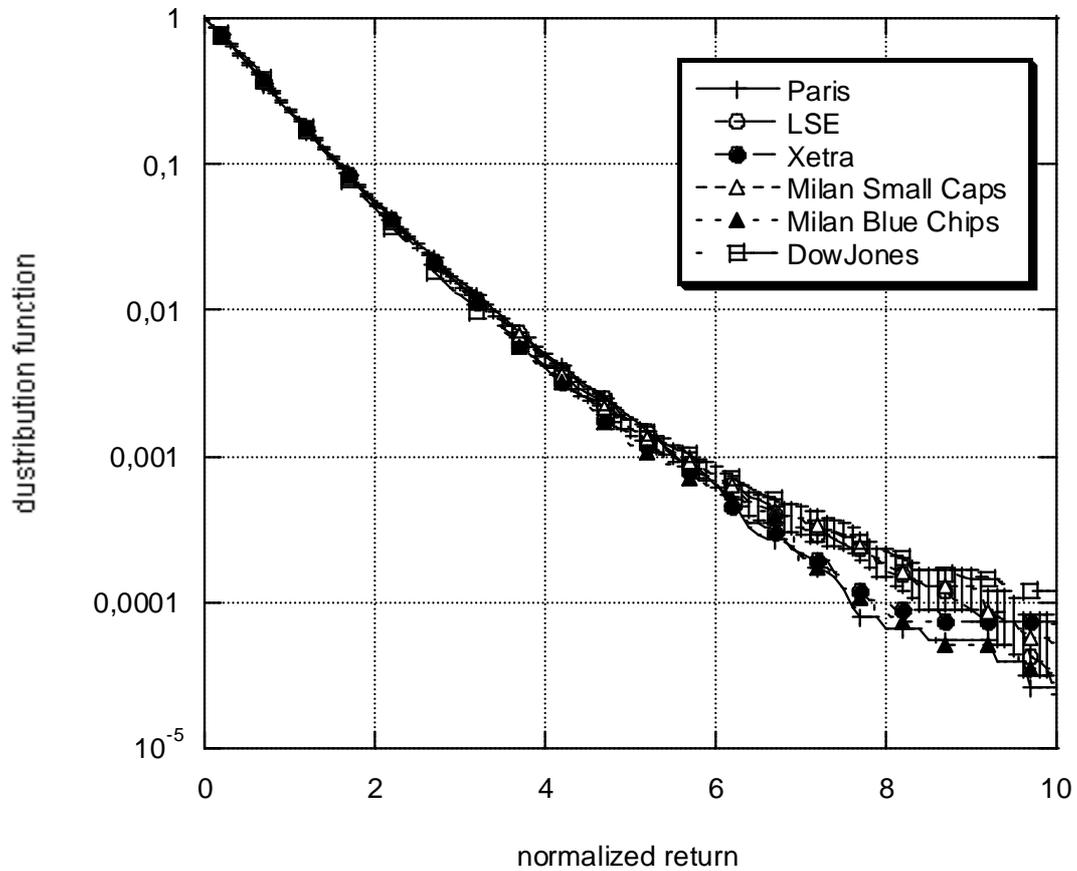

Fig. 2

Log plot of the average cumulative distribution function of absolute daily returns of stocks traded on different markets. For a description of each database see table I. A very good overlap is observed up to a normalized return of 5 covering about 99,9% of the available data. On the abscissa the return is normalized to its standard deviation of each database.

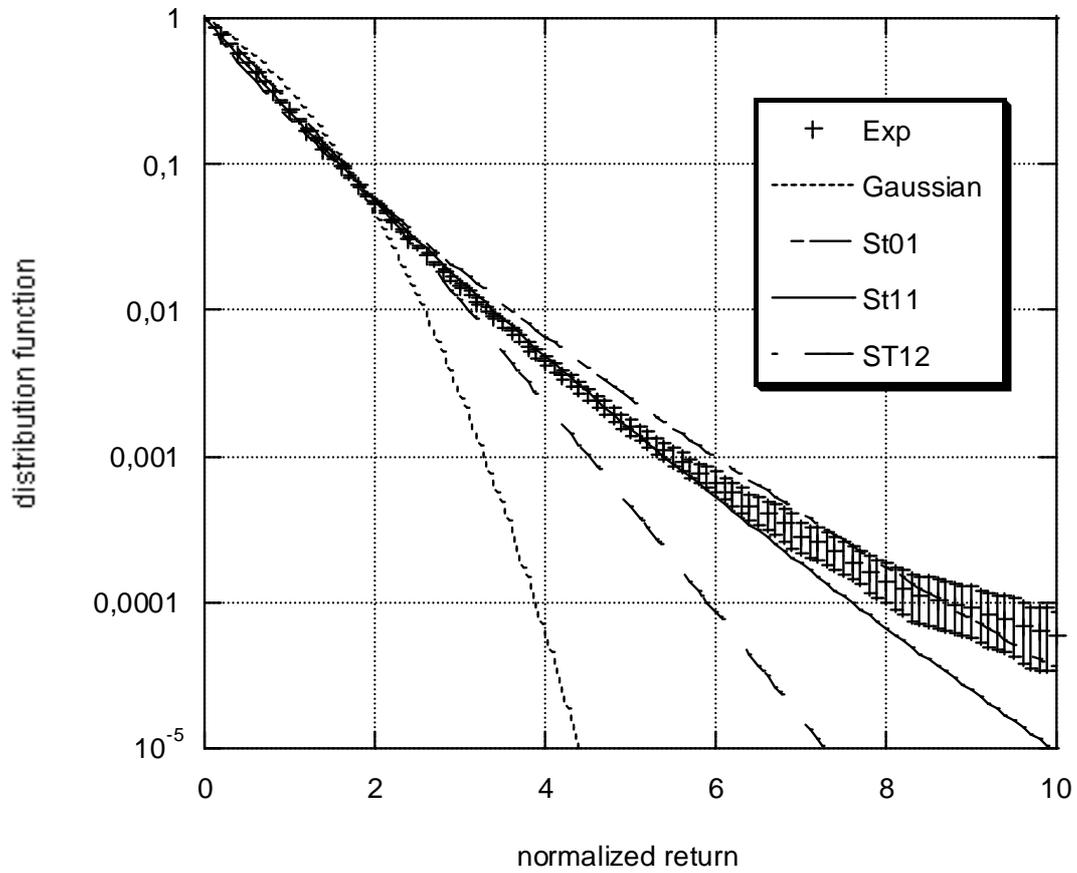

Fig. 3

This figure shows the average cumulative distribution function of stock price absolute returns including data of all daily dataset listed in table I. Error bars are constructed as the standard deviation of the average. The ST11 model described in the text reproduces the observations up to a normalized return of about 5. A Gaussian distribution is also shown for comparison.

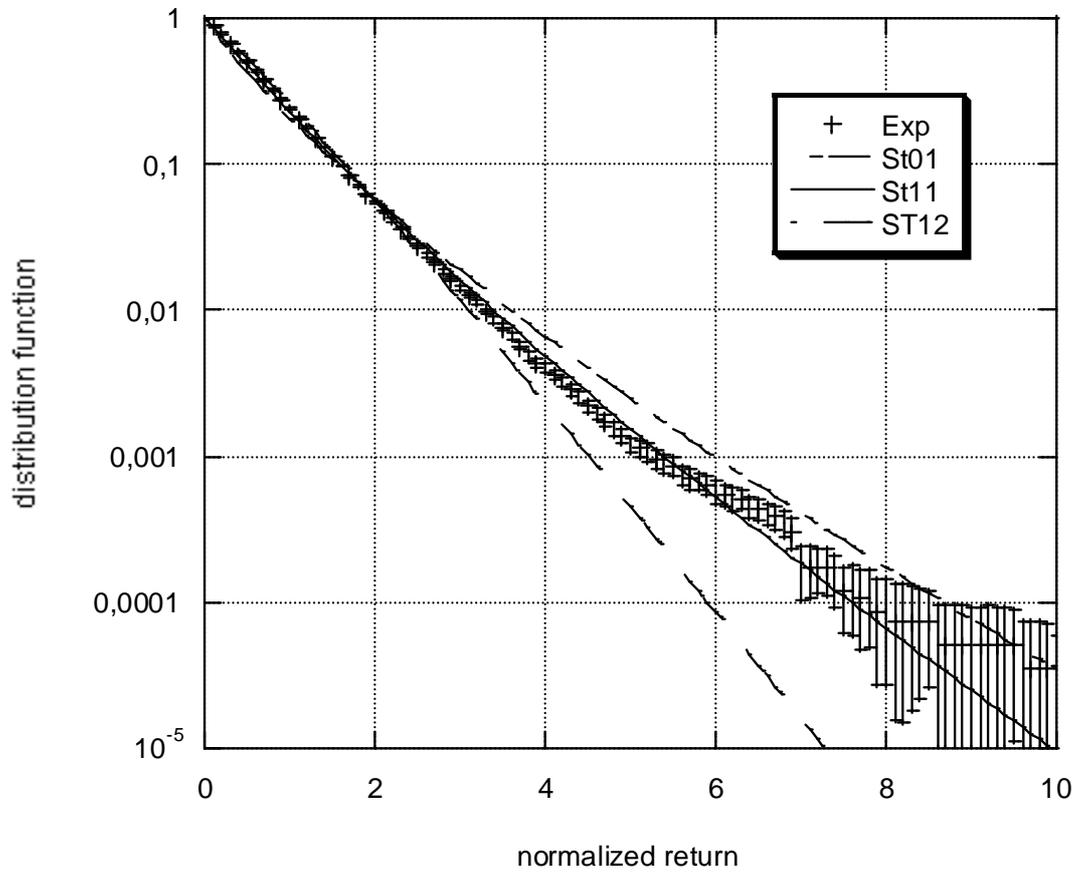

Fig. 4

This figure shows the average cumulative distribution function of stock price two hours absolute returns using data of 13 stocks included in the last dataset listed in table I. Error bars are constructed as the standard deviation of the average. Here again the ST11 model described in the text provides the best match to the observations.

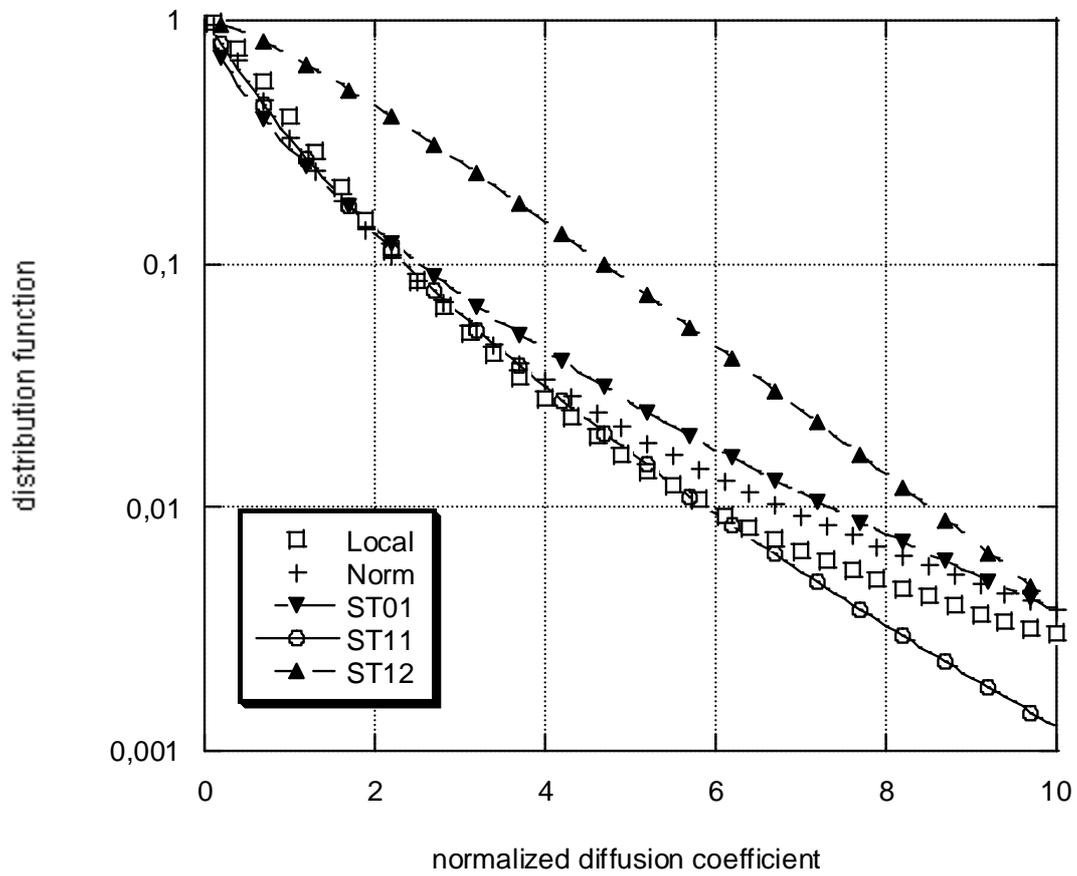

Fig. 5

Crosses represent diffusion coefficients normalized to their average over the whole database while open squares are obtained normalizing over the average value of the previous three days. Continuous line represents the prediction of the ST11 model and accounts satisfactorily for the experimental data.

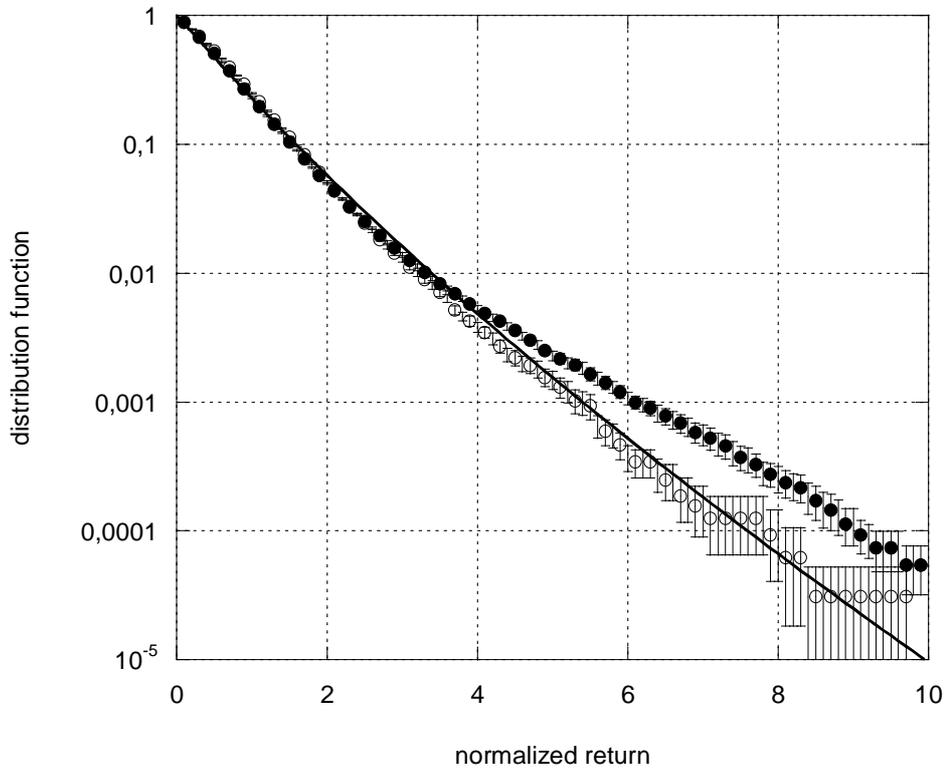

Fig. 6

In this plot we analyse Dow Jones components from January 1990 to April 2010. Full dots represent daily absolute return data points for the whole dataset while open dots describe a subset of data consisting of days when the volatility VIX index is comprised between 15% and 25%. The continuous line is the cumulative distribution predicted by the ST11 model. Tails are very prominent for the complete dataset but are well described by our model when the VIX index is constrained.